\PassOptionsToPackage{unicode}{hyperref}
\PassOptionsToPackage{hyphens}{url}
\documentclass[
  11pt,
]{article}
\usepackage{xcolor}
\usepackage[margin=1in]{geometry}
\usepackage{amsmath,amssymb}
\usepackage{iftex}
\ifPDFTeX
  \usepackage[T1]{fontenc}
  \usepackage[utf8]{inputenc}
  \usepackage{textcomp} 
\else 
  \usepackage{unicode-math} 
  \defaultfontfeatures{Scale=MatchLowercase}
  \defaultfontfeatures[\rmfamily]{Ligatures=TeX,Scale=1}
\fi
\usepackage{lmodern}
\ifPDFTeX\else
\fi
\IfFileExists{upquote.sty}{\usepackage{upquote}}{}
\IfFileExists{microtype.sty}{
  \usepackage[]{microtype}
  \UseMicrotypeSet[protrusion]{basicmath} 
}{}
\makeatletter
\@ifundefined{KOMAClassName}{
  \IfFileExists{parskip.sty}{%
    \usepackage{parskip}
  }{
    \setlength{\parindent}{0pt}
    \setlength{\parskip}{6pt plus 2pt minus 1pt}}
}{
  \KOMAoptions{parskip=half}}
\makeatother
\NewDocumentCommand\citeproctext{}{}

\makeatletter
 \let\@cite@ofmt\@firstofone
 \def\@biblabel#1{}
 \def\@cite#1#2{{#1\if@tempswa , #2\fi}}
\makeatother
\newlength{\cslhangindent}
\newlength{\csllabelwidth}
\newenvironment{CSLReferences}[2] 
 {\begin{list}{}{%
  \setlength{\itemindent}{0pt}
  \setlength{\leftmargin}{0pt}
  \setlength{\parsep}{0pt}
  \ifodd #1
   \setlength{\leftmargin}{\cslhangindent}
   \setlength{\itemindent}{-1\cslhangindent}
  \fi
  \setlength{\itemsep}{#2\baselineskip}}}
 {\end{list}}
\usepackage{calc}

\newcommand{\CSLLeftMargin}[1]{\parbox[t]{\csllabelwidth}{\strut#1\strut}}
\newcommand{\CSLRightInline}[1]{\parbox[t]{\linewidth - \csllabelwidth}{\strut#1\strut}}

\usepackage{bookmark}
\IfFileExists{xurl.sty}{\usepackage{xurl}}{} 
\makeatletter
\@ifundefined{xmpquote}{}{}
\makeatother
\hypersetup{
  pdftitle={Who is scientific code for? Maintaining human-readable landmarks in agent-written code},
  hidelinks,
  pdfcreator={LaTeX via pandoc}}

\title{Who is scientific code for? Maintaining human-readable landmarks
in agent-written code}
\usepackage{etoolbox}
\makeatletter
\providecommand{\subtitle}[1]{
  \apptocmd{\@title}{\par {\large #1 \par}}{}{}
}
\makeatother
\subtitle{Elle O'Brien, University of Michigan School of Information}
\author{}
\date{}

\begin{document}
\maketitle

A longstanding assumption of scientific research involving code is that
at least one person understands why code exists {[}6{]}. With increasing
use of agents, this assumption no longer holds. For many kinds of
research, sufficient understanding is still essential---code must
correctly reflect scientific choices and assumptions. Storey {[}9{]} has
recently proposed that agent-involved software work accelerates the
accumulation of cognitive debt (erosion of shared understanding) and
intent debt (failure to capture goals in legible artifacts). I study how
scientists program, and I'm seeing fascinating responses to the debts
Storey articulated: scientists are inventing their own idiosyncratic
conventions for marking which parts of a codebase are meant as
human-readable landmarks and which are context for agents. That means
intent is being captured through customs that are never made explicit or
shared, and this may be quietly making scientific software
infrastructure less amenable for collaboration and reuse.

\subsection{What I'm seeing}\label{what-im-seeing}

My observations come from an ongoing contextual inquiry: I watch
scientific programmers (both scientists and research software engineers)
work with agentic tools on their own research software, in their own
environments, and interview them as they go. I have completed four cases
so far.

The scientists I've met are grappling with a new kind of work: in
addition to making sure agent-produced code is reasonably correct and
consonant with their intentions, they need to rein in the
\emph{magnitude} of code complexity it is possible to produce, which
exceeds the capacity for any one person to review and understand. For
many scientists (\emph{terminology note:} for simplicity, I will use the
term ``scientist'' for any person working in a scientific context,
including research software engineers), oversight was already a limited
resource---code review is not widely practiced {[}1, 7{]} and many code
projects have a small number of contributors {[}2, 5, 7{]}. To manage
this pressure, I observe scientific programmers intentionally creating
\emph{landmarks} for human understanding in a stream of agent-produced
artifacts.

What does a landmark look like? So far, it appears quite personal. One
research software engineer I observed created a mental rule: their agent
can produce anything locally, but what gets committed must reflect the
version the scientist accepts, and the commit message is typed by the
scientist like a journal entry (what they might make in a lab notebook).
The commit message marks their ``why'' for changing the code and is what
they would return to a few weeks later as a document of their choices.

Here's another approach: a post-doctoral researcher allows the agent to
commit again and again on a new branch and author commit messages and
pull requests on GitHub. But only what gets \emph{merged}, the diff, is
meant to be understood by a human as the choice. When the scientist
wants to understand what choices he made in the past and why, he looks
at the \emph{diffs} that have been accepted into main. He therefore
strives for each diff to be minimal and uncomplicated, squashing
irrelevant commits so each merge is self-explanatory at a glance.
Meanwhile, the instructions he gives the agent (the specs) are transient
and overwritten constantly, never committed to the repository or seen by
another person. Even though spec-recording markdown files are in theory
a way of making intent explicit and visible, this scientist prefers code
as a reference.

Yet another uses a plugin for spec-driven development, and writes spec
documents to markdown. These go through rounds of iteration---a
brainstorm, then a draft, then a final spec, which is committed to the
repository alongside changes. Code is not reviewed, the specs are.

What if there is no version control? This is how many scientists work
{[}3, 7{]}, particularly during experimental coding, which exists to
interact with the data and discover its patterns and anomalies {[}4,
10{]}; some scientists live in this mode; others toggle between it and a
production mode. One scientist mentions that now that writing code is
quick, he spends most of his time exploring alternative ways to analyze
his data---what if he changes the way a variable is represented
mathematically? Does that make the downstream analyses easier to
understand? For this scientist, who does not use Git version control for
code, he maintains two artifacts: first, the chat logs record every
choice explored; second, all code ever produced goes into a single
script. Because context is finite, he intentionally narrows down which
analyses the agent writes into the file. The code is not just full of
analyses, but also comments left by the agent explaining what they do
and why. He doesn't read them, but he leaves them in as context for the
agents in the future. So the code keeps growing as he appends more
alternatives and asides for the agents, rarely deleting. And if he does
need to revert to an old way of doing things, he relies on the chat
memory to reconstruct it (re-executing is cheap).

This scientist knew Git---he used it for manuscript tracking in
Overleaf---so this isn't a simple matter of awareness or education.
Watching him work reminded me of a dilemma I recently faced in my own
data analysis coding.

During analysis, I often try several alternative ways to model data to
confirm any findings are robust {[}8{]}---if switching a small detail of
a statistical modeling choice changes the overall pattern, I want to
know this. I was delighted to find that an agent could quickly swap out
different choices in the code (family of statistical models,
thresholding choices), rerun the entire pipeline from scratch, and
detect any substantial drift in results that occurred. But I was
paralyzed when it came to handling my experiments in Git, because I view
the purpose of a repository to ultimately be reproducible and readable,
and this kind of multiverse code feels like unnecessary complexity to
foist on a potential future reader (particularly because it exists to
confirm that these choices \emph{don't} matter to the end results of the
analysis!).

In theory, I want Git history to reflect these explorations, so they are
documented and agents can reference the attempted paths. Yet committing
experimental code chunks that won't be developed further, or maintaining
parallel branches for dead ends, conflicts with how I understand the
purpose of branches or commits. I settled into committing haphazardly
and without any real documentation of what had been tried, which was
represented only partially in my agent logs and my own memory.

In other kinds of scientific coding, I have no problems using Git with
an agent---but in this particular kind of data analysis work, my mental
model broke. I would bet this is not uncommon, but I don't have a full
theory yet of why and when this occurs.

To summarize: a landmarking strategy is a scientist's own rules for how
to triage artifacts for review and record decisions about what to build
(or not to build). When scientists use version control, they may
repurpose it for maintaining the place where the \emph{why} of a code
choice lives. This can render previous uses of version control systems
overloaded or obsolete---for example, we can no longer assume that
commits are intended to be reviewable units, or that commit messages are
for people to read.

\subsection{Collaboration}\label{collaboration}

All of these scientists have collaborators who have varying degrees of
contact with their code. Because each landmarking strategy is
individually determined, conventions fragment across teams. While I do
not have direct evidence to corroborate this yet, I expect that this
fragmentation will complicate collaboration. Where should a newcomer to
a codebase---whether a contributor, a reader who wishes to replicate a
published result, or a new graduate student picking up work left behind
by a previous lab member---look to establish their understanding?
Importantly, understanding might not encompass only the \emph{choices}
that have been made in the code to date, but also the elements where
cognitive debt has never been paid down: presumably parts of the code
that show evidence of strong oversight, whether through a review process
or testing, should warrant more trust than those which were produced
under time pressure with little validation.

The strategies I have observed are not necessarily interoperable with
how other scientists will work. Consider that many scientists don't
appear to be using agents at all, but regular chatbots. In our 2025
survey of \textgreater800 scientific programmers, roughly three quarters
of adopters worked through general-purpose conversational tools like
ChatGPT rather than developer tools {[}7{]}. Imagine that someone who
maintains clean Git diffs as their cognitive landmark is collaborating
with someone who doesn't work with an agent, but a chatbot that cannot
interface with Git directly. If they upload the current version of the
code, the chatbot is unlikely to surface that the version diffs or
commit messages are meaningful touchpoints for understanding.
Intriguingly, some efforts are underway to build systems that help
non-Git users who code with agents access shared context represented in
repository artifacts like GitHub issues {[}11{]}--this direction could
surface some design principles for scientific teams with highly
heterogeneous software practices.

Even within a single snapshot of code isolated from versioning, there
can be substantial context fragmentation. The scientist making a long
analytical script sends it to another scientist, whose agent has access
to any comments left behind by the author's agent, but not the full
conversational transcript that produced it (which serves as the version
control and intent log). Is the collaborator scientist supposed to read
the in-line comments? Would they feel differently about how to use them
if they knew the author had never read them himself, and only left them
behind for other agents?

This redefinition of what is human-readable versus agent-readable might
also create incompatibilities across resource differentials: for a
scientist with an effectively unlimited token budget, what an agent can
recover about a collaborator's codebase and its history may differ
sharply from what a token-poor scientist can recover.

For these reasons, while I think agents have the potential to
\emph{improve} the interoperability and sharability of scientific code
in the abstract, what I am seeing so far is that the idiosyncrasies of
scientific code may in fact be growing more particular---a necessity,
given the multiverses of analytical possibilities and software features
that are now cheap to produce. I therefore expect that teams that work
together to explicitly delineate what artifacts are meant to be
human-readable, versus context for agents, will be better able to
develop, document, and maintain scientific code bases in the coming
years.

\subsection{About the author}\label{about-the-author}

Elle O'Brien is a lecturer and researcher at the University of Michigan
School of Information, where she studies how scientists program with
generative AI programming tools. She also serves as the director of the
Graduate Data Science Certificate program at the Michigan Institute for
Data, AI and Society. O'Brien's research is generously supported by the
Alfred P. Sloan Foundation's Digital Technology Fund and Schmidt
Sciences.

\subsection{References}\label{references}

\protect\phantomsection\label{refs}


\begin{CSLReferences}{0}{0}
\bibitem[\citeproctext]{ref-Carver2022AStates}
\CSLLeftMargin{{[}1{]} }%
\CSLRightInline{Carver, J.C. et al. 2022. {A survey of the state of the
practice for research software in the United States}. \emph{PeerJ
Computer Science}.
https://doi.org/\href{https://doi.org/10.7717/peerj-cs.963}{10.7717/peerj-cs.963}.}

\bibitem[\citeproctext]{ref-Howison2011ScientificCollaboration}
\CSLLeftMargin{{[}2{]} }%
\CSLRightInline{Howison, J. and Herbsleb, J.D. 2011. {Scientific
software production: Incentives and collaboration}. \emph{Proceedings of
the ACM conference on computer supported cooperative work, CSCW}.
https://doi.org/\href{https://doi.org/10.1145/1958824.1958904}{10.1145/1958824.1958904}.}

\bibitem[\citeproctext]{ref-Jay2022NotThem}
\CSLLeftMargin{{[}3{]} }%
\CSLRightInline{Jay, C. et al. 2022. \emph{{"Not everyone can use Git":
Research Software Engineers' recommendations for scientist-centred
software support (and what researchers really think of them)}}.
https://doi.org/\href{https://doi.org/10.48420/17313215.v1}{10.48420/17313215.v1}.}

\bibitem[\citeproctext]{ref-Kery2017ExploringProgramming}
\CSLLeftMargin{{[}4{]} }%
\CSLRightInline{Kery, M.B. and Myers, B.A. 2017. {Exploring exploratory
programming}. \emph{Proceedings of IEEE symposium on visual languages
and human-centric computing, VL/HCC}.
https://doi.org/\href{https://doi.org/10.1109/VLHCC.2017.8103446}{10.1109/VLHCC.2017.8103446}.}

\bibitem[\citeproctext]{ref-Milewicz2019CharacterizingProjects}
\CSLLeftMargin{{[}5{]} }%
\CSLRightInline{Milewicz, R. et al. 2019. {Characterizing the roles of
contributors in open-source scientific software projects}. \emph{IEEE
international working conference on mining software repositories}.
https://doi.org/\href{https://doi.org/10.1109/MSR.2019.00069}{10.1109/MSR.2019.00069}.}

\bibitem[\citeproctext]{ref-Naur1985ProgrammingBuilding}
\CSLLeftMargin{{[}6{]} }%
\CSLRightInline{Naur, P. 1985. {Programming as theory building}.
\emph{Microprocessing and Microprogramming}.
https://doi.org/\href{https://doi.org/10.1016/0165-6074(85)90032-8}{10.1016/0165-6074(85)90032-8}.}

\bibitem[\citeproctext]{ref-OBrien2025AProgram}
\CSLLeftMargin{{[}7{]} }%
\CSLRightInline{O'Brien, G. et al. 2025. {A survey of generative AI
adoption and perceived productivity among scientists who program}.
https://doi.org/\href{https://doi.org/10.48550/arXiv.2512.19644}{10.48550/arXiv.2512.19644}.}

\bibitem[\citeproctext]{ref-Sarma2023Multiverse:Notebooks}
\CSLLeftMargin{{[}8{]} }%
\CSLRightInline{Sarma, A. et al. 2023. {multiverse: Multiplexing
Alternative Data Analyses in R Notebooks}. \emph{Conference on Human
Factors in Computing Systems - Proceedings}.
https://doi.org/\href{https://doi.org/10.1145/3544548.3580726}{10.1145/3544548.3580726}.}

\bibitem[\citeproctext]{ref-Storey2026FromAI}
\CSLLeftMargin{{[}9{]} }%
\CSLRightInline{Storey, M.-A. 2026. {From Technical Debt to Cognitive
and Intent Debt: Rethinking Software Health in the Age of AI}.
https://doi.org/\href{https://doi.org/10.48550/arXiv.2603.22106}{10.48550/arXiv.2603.22106}.}

\bibitem[\citeproctext]{ref-Sutherland-Keller2025ResearchCosmology}
\CSLLeftMargin{{[}10{]} }%
\CSLRightInline{Sutherland-Keller, W. 2025. \emph{{Research Software
Systems: Exploration and Infrastructure in Observational Cosmology}}.}

\bibitem[\citeproctext]{ref-Udell2026VibeSport}
\CSLLeftMargin{{[}11{]} }%
\CSLRightInline{Udell, J. 2026. {Vibe coding as a team sport}.
\url{https://blog.jonudell.net/2026/06/17/vibe-coding-as-a-team-sport/}.}

\end{CSLReferences}
\end{document}